# Uncertainty Management in Power System Operation Decision Making

Mohammad Hemmati[1], Behnam Mohammadi-Ivatloo[1], Alireza Soroudi[2]

[1] Faculty of Electrical and Computer Engineering, University of Tabriz, Tabriz, Iran

[2] University College Dublin, Dublin, Ireland

## I. Abstract

Due to the penetration of renewable energy resources and load deviation, uncertainty handling is one of the main challenges for the power system. Therefore, the need for accurate decision making in a power system under the penetration of uncertainties is essential. However, decision-makers should use suitable methods for uncertainty management. In this chapter, some of uncertainty modeling methods in power system studies are analyzed. At first, multiple uncertain parameters which the power system deals with them are introduced, then, some useful uncertainty modeling methods are introduced. To show the uncertainty modeling process and its effect on the decision making, a microgrid consisting multiple uncertain parameters is considered, and stochastic scenario-based approach is used for uncertainty modeling. The scheduling of microgrid in presence of different types of uncertainty is solved from the profit maximization point of view. |The simulation results are presented for 33-bus microgrid which shows the effectiveness of the proposed method for decision making under high level of uncertainty.

## II. Introduction

The development of power system and the emergence of new energy concepts such as microgrids and smart grids have caused various challenges in the scheduling and operation of these networks. Modern power system scheduling like conventional networks can be performed for short, medium and long-term periods [1, 2]. However, the need for accurate decision making for these periods is essential. Management the challenges of the power system is a set of decision problem affiliated to a different part (e.g., scheduling, investment, and operation) where decision makers must distinguish all alternatives from cost, revenue or risk point of views.

Besides the all of modern power system challenges, the growth of total installed renewable energy resources with probabilistic nature cause the complex planning of power system. It is forecasted that the renewable energy share can reach 36% by 2030 [3]. This massive generation power by renewable energy is associated with uncertainty. Renewable energy resource power is depended on initial sources like wind and solar. Dependence of these resources to climate condition causes the increasing of uncertainty in generated power [4]. However, as renewable energy penetration increases, there will be an increase in the uncertainty associated with the power system. Hence, uncertainty modeling and suitable addressing in planning and operation of a power system is essential [5, 6].

The uncertainty handling is one of the main issues of decision makers in power system [7]. All of the uncertain parameters that the power systems faced with them can be classified into economical and technical parameters. According to [7], technical uncertain parameters are related to the topological of network like failure rate of transmission lines or generators and etc. Another technical uncertain parameter which is effected on the operating decisions, are generation and demand value in system. The economical uncertain parameters contain energy price, economic growth, environmental policies and etc., which faces the decision making process with multiple challenges [8]. Table I categorizes all of the possible uncertain parameters in the power system that should be handled for suitable operation of the system.

Table I. Classified uncertain parameters in power system [7].

| Technical parameters | | Economical parameters |
|---|---|---|
| **Operational parameters** | **Topological parameters** | |
| Load demand | Line outage | Economic growth |
| Generation output | Generator outage | Price levels |
| | | Governmental regulation |
| | | Unemployment rate |
| | | Fuel price |

However, in scheduling and operation of the power system, the main objective is cost minimization or profit maximization [9]. Therefore, the main uncertain parameters which we are facing with them includes generated power of renewable energies, load demand of consumption and energy price. So, we focus on the modeling of these parameters. In the following section, a comprehensive review on uncertainty handling in power system will be provided. Then, we will choose one of them and provide the example to describe the uncertainty modeling in power system.

## III. Uncertainty management in power system: a review

There are multiple methods for handling of uncertainty in power systems. The main feature that causes the discrepancy between multiple methods is in line with the different technique used to describe the uncertain input data and parameters [10] and the degree of uncertainty. In the following section, some of the main uncertainty handling approaches will be described.

1. **Probabilistic method**

One of the simple and earliest methods which assumes all of the input uncertain parameters as random variables, is a probabilistic approach [11]. In this method, each variable has a known probability density function (PDF) and the uncertain parameters are modeled according to the correspondingly PDF. The probabilistic programming was first introduced by Dantzing [12].

There are three different probabilistic approach techniques which are used for uncertainty management: Monte-Carlo simulation (MCS), Point estimation method (PEM) and scenario-based optimization method [13].

In MCS, $n_s$ samples for each uncertain parameter are generated according to the corresponding PDF of each one. Assuming $n_s = \{n_1, n_2, ....., n_s\}$, $y^s = f(n_s)$ is calculated. The process is repeated for a lot of iterations, until the average value of each uncertain parameter is obtained. Some of the valuable works that use MCS in probabilistic programming can be found in [13-17].

The PEM is based on the concept of statistical moments of input uncertain parameters. Unlike the MCS, PEM only generates two samples for each parameter. Some of the valuable works that use MCS in probabilistic programming can be found in [18].

The decision making by scenario-based is another technique based on probability theory. A list of scenarios is generated using the corresponding PDF of each uncertain parameter. In section 3, an example of modern power system and scenario-based optimization approach is implemented to uncertainty modeling. The details of this technique are presented [19]. Some of the valuable works that use scenario-based decision-making method can be found in [20].

## 2. Information Gap Decision Theory

In some cases, the uncertain parameters do not follow a PDF or the PDF is not known by the decision maker. In such cases, the information gap decision theory (IGDT) is used to model the uncertainty [7, 21]. In IGDT, the robustness is defined as the inviolability of euphoria of a predefined constraint [21]. Suppose $X$ is the set of uncertain parameters and all of the components of this set is equal to its predicated value ($X = \bar{X}$), so the predicated value for objective function ($\bar{y}$) is obtained. When the value of the uncertain parameter is not equal to the predicated value and is unknown, IGDT is implemented to find a good solution which gives the robustness feature to the value of objective function against the predicting error. Some of the valuable works that use IGDT can be found in [22-26].

### 3. Robust optimization

The robust optimization technique is another important uncertainty modeling tool in the power system studies. In this method it is assumed that the uncertain parameter belongs to an uncertainty set and it is tried to make the optimal decision considering this fact. In other words, the decision variables are found in a way that the objective function remains optimal even if the uncertain parameter takes its worst-case value [27, 28]. For example, consider a power system which includes wind and solar sources and load demand are all associated with uncertainty. The robust optimization first finds the worst case for the realization of uncertain parameters and then find the optimal scheduling of system flexibilities. Using this method makes the system robust against the uncertain parameters and the value of objective function remains reasonable even if the uncertain parameters do not take their predicted values. In this way, the risk of decision in scheduling and operation of system can be controlled [29]. Some of the valuable works that use robust optimization method can be found in [30, 31].

## IV. Problem formulation

In this section, an example of modern power system scheduling problem is presented which involves the uncertainty. The microgrid concept as a futuristic distribution network is one of the modern power systems that has attracted much interest in recent years. Each microgrid includes multiple components: distributed generation (generation units and energy storage system), different loads (controllable and non-controllable load), switches and etc. The penetration of renewable energy like PV and wind in the microgrid with probabilistic nature, may cause the complexity for the scheduling and operation of these networks. Fig. 1 shows the 33-bus microgrid consists of the wind and PV units as renewable energy resources, along with the dispatchable units (diesel generators and energy storage system), which can exchange the power with the upstream network.

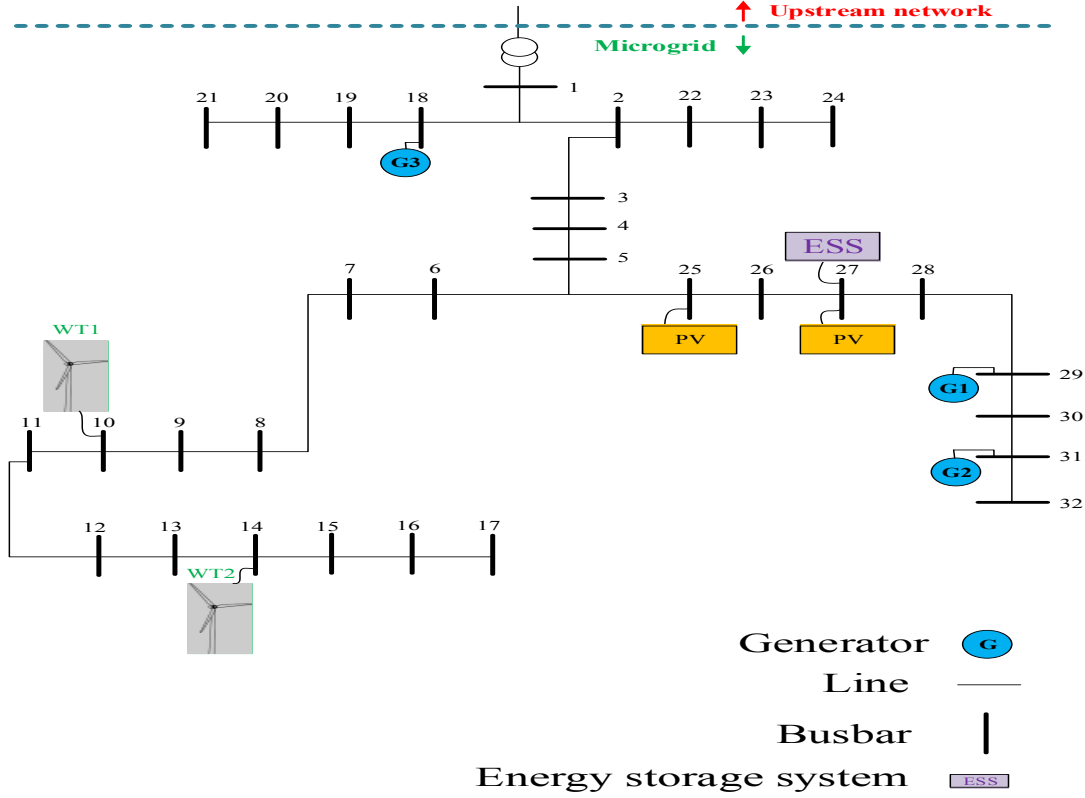

**Fig. 1: Schematic of microgrid with multiple units [32].**

In this problem, the main objective is profit maximization, which is obtained by the revenues minus the system costs. The objective function can be formulated as follows:

$$Maximize \quad Profit = RV - TC \tag{1}$$

where $RV$ is MG's revenue (\$), $TC$ is total cost (\$) of MG. The revenue for MG's is calculated according to:

$$RV = \sum_{\omega=1}^{N_\omega} \pi_\omega \sum_{t=1}^{T} \sum_{l=1}^{L} \lambda_t^L \left( P_{l,t,\omega} \right) + r \sum_{t=1}^{T} \lambda_t^N P_{t,\omega}^N \qquad \omega = 1, 2, ..., N_\omega \tag{2}$$

where $\lambda_t^L$ is power market price that the consumers pay at $t^{th}$ time, $P_{l,t,\omega}$ is active power demand of $l^{th}$ load at $t^{th}$ time in $\omega^{th}$ scenario, $P_{t,\omega}^N$ is power sold to the upstream network at $t^{th}$ time, $\lambda_t^N$ is price of selling/purchasing power with the upstream network at $t^{th}$ time in $\omega^{th}$ scenario, r is the binary variable that separates selling or purchasing state; $r$ is one if MG sells power to the upstream network, $t$ is the index of time, $i$ is index of dispatchable unit, $l$ is index of load, $\omega$ is index of the scenario and $N_\omega$ is the number of scenarios.

The total operation cost of MG includes fuel cost, startup and shutdown cost, emission cost of the dispatchable units and the cost of purchasing power from the upstream network that is calculated as follows:

$$TC = \sum_{\omega=1}^{N_\Omega} \pi_\omega \sum_{i=1}^{N} \sum_{t=1}^{T} (F(P_{i,t,\omega})X_{i,t,\omega} + SU_{i,t,\omega} + SD_{i,t,\omega} + (C^{emi}.\tau_i.P_{i,t,\omega})) + (1-r)\sum_{t=1}^{T} \lambda_t^N P_{t,\omega}^b \quad (3)$$

where $\pi_\omega$ is the probability of $\omega^{th}$ scenario and $N_\Omega$ is the number of total scenarios. $F(P_{i,t})$ is fuel cost consumption function of $i^{th}$ generator at $t^{th}$ time that is calculated as follows:

$$F(P_{i,t}) = a_i + b_i P_{i,t} + c_i (P_{i,t})^2 \quad i \in N, t \in T \quad (4)$$

In (3), $a_i$, $b_i$ and $c_i$ are cost coefficients of $i^{th}$ generator, $P_{i,t,\omega}$ is the power output of $i^{th}$ generator at $t^{th}$ time and $\omega^{th}$ scenario. $X_{i,t,\omega}$ is the commitment state of $i^{th}$ generator at $t^{th}$ time and $\omega^{th}$ scenario.

The second and the third terms of (2), show the technical cost of generator named: startup and shutdown cost, respectively. The fourth term represents the emission cost of $i^{th}$ generator, $\tau_i$ is the emission factor *(kg/kWh)* of $i^{th}$ generator and $C^{emi}$ is emission cost of $i^{th}$ generator *($/kg)* [33]. The fifth term represents the cost of purchased power and $P_{t,\omega}^b$ is purchased power from the upstream network at $t^{th}$ time and $\omega^{th}$ scenario.

1. **Constraints**

- *Power balance constraint*

The sum of generated power by dispatchable (generators and ESS) and non-dispatchable (PV and WT) units and exchanged power with the upstream network must be greater or equal to the sum of forecasted power demand and power loss as follows:

$$\sum_{i=1}^{N} P_{i,t,\omega} X_{i,t,\omega} \pm \sum_{e=1}^{E} P_{ESS_{e,t,\omega}} X_{e,t,\omega} + \sum_{k=1}^{K} P_{PV_{k,t,\omega}} + \sum_{w=1}^{W} P_{WT_{w,t,\omega}} \pm P_{grid_{t,\omega}} \geq P_{loss_{t,\omega}} + \sum_{l=1}^{L} P_{l,t,\omega} \quad \begin{array}{c} t \in T \\ \omega \in N_{\omega} \end{array} \quad (5)$$

where $P_{ESS_{e,t,\omega}}$ is charged/discharged power by $e^{th}$ ESS at $t^{th}$ time in $\omega^{th}$ scenario, $X_{e,t,\omega}$ is commitment state of $e^{th}$ ESS at $t^{th}$ time in $\omega^{th}$ scenario. $X_{e,t,\omega}^{c}$ and $X_{e,t,\omega}^{d}$ are binary variables denoting the $e^{th}$ ESS charging and discharging mode at $t^{th}$ time in $\omega^{th}$ scenario, respectively. $P_{pv_{k,t,\omega}}$ is power output of $k^{th}$ PV at $t^{th}$ time in $\omega^{th}$ scenario, $P_{WT_{w,t,\omega}}$ is power output of $w^{th}$ WT at $t^{th}$ time in $\omega^{th}$ scenario $P_{grid_{t,\omega}}$ is exchanged power with the upstream network at $t^{th}$ time and $\omega^{th}$ scenario, if $r$ is one, $P_{grid_{t,\omega}} = P_{t,\omega}^{N}$ and power sold to the upstream network, otherwise, $P_{grid_{t,\omega}} = P_{t,\omega}^{b}$ and power purchased from the upstream network. $P_{loss_{t,\omega}}$ is total power loss at $t^{th}$ time in $\omega^{th}$ scenario.

- *Inequality constraints*

The generated power by generators and exchanged power are bounded by upper and lower limits as follows:

$$P_{i}^{\min}(t) \leq P_{i}(t) \leq P_{i}^{\max}(t) \quad t \in T \quad (6)$$

$$P_{grid}^{\min}(t) \leq P_{grid}(t) \leq P_{grid}^{\max}(t) \quad t \in T \quad (7)$$

where $P_i^{\min}$ and $P_i^{\max}$ are minimum and maximum power output of $i^{th}$ generator, $P_{grid}^{\min}$ and $P_{grid}^{\max}$ are minimum and maximum exchanged power with the upstream network.

- *Energy storage constraints*

The power that charged or discharged by ESS is bounded by the upper and lower limits:

$$P_{dis_e}^{\min}(t) X_{e,t}^d \leq P_{ESS_e}(t) \leq P_{dis_e}^{\max}(t) X_{e,t}^d \qquad \begin{array}{c} \forall t \in T \\ \forall e \in E \end{array} \qquad (8)$$

$$P_{char_e}^{\min}(t) X_{e,t}^c \leq P_{ESS_e}(t) \leq P_{char_e}^{\max}(t) X_{e,t}^c \qquad \begin{array}{c} \forall t \in T \\ \forall e \in E \end{array} \qquad (9)$$

where $P_{dis}^{\min}$ and $P_{char}^{\min}$ are minimum power discharged and charged by $e^{th}$ ESS, respectively. $P_{dis}^{\max}$ and $P_{char}^{\max}$ are maximum power discharged and charged by $e^{th}$ ESS, respectively. $X_{e,t}^d$ and $X_{e,t}^c$ are binary variables for discharging and charging modes of $e^{th}$ ESS, when charging, the charging state $X_e^c$ is one and discharging state $X_e^d$ is zero, so the minimum and maximum limits of charging mode is imposed (9). When discharging, the discharging state $X_e^d$ is one and charging state $X_e^c$ is zero, so the minimum and maximum limit of discharging is imposed (8). To separate the ESS operation modes (ESS cannot operate in both charging and discharging modes, simultaneously), another constraint is considered as follows:

$$X_e^c + X_e^d \leq 1 \qquad \forall e \in E \qquad (10)$$

Energy storage system state of charge (SOC) constraint is calculated as:

$$0 \leq SOC_{e,t} \leq SOC_e^{\max} \qquad \forall e \in E \qquad (11)$$

where $SOC_e^{\max}$ is maximum state of charge of $e^{th}$ ESS.

The energy storage system is subject to minimum charging and discharging time limits, respectively [34]:

$$T_e^{char} \geq MC_e \left( X_{e,t}^c - X_{e,t-1}^c \right)$$
$$T_e^{dis} \geq MD_e \left( X_{e,t}^d - X_{e,t-1}^d \right) \qquad \forall e \in E \;,\; \forall t \in T \qquad (12)$$

where $MC_e$ and $MD_e$ are minimum charging and discharging time of $e^{th}$ ESS, respectively. $T_e^{char}$ and $T_e^{dis}$ are the number of continuous charging and discharging time (per hour) of $e^{th}$ ESS.

- *Minimum up/down time constraint:*

As we know, a dispatchable unit is limited by minimum up and down times. The time during which a unit must be on/off after being startup/shutdown, called minimum up/down time constraints, respectively, as following [35]:

$$MUT_i \left( X_{i,t} - X_{i,t-1} \right) \leq T_i^{on}$$
$$MDT_i \left( X_{i,t-1} - X_{i,t} \right) \leq T_i^{off} \qquad (13)$$

where $MUT_i$ and $MDT$ are minimum up and down time of $i^{th}$ generator, $X_{i,t}$ and $X_{i,t-1}$ are commitment status of $i^{th}$ generator at $t$-$1^{th}$ and $t^{th}$ time, respectively. $T_i^{on}$ and $T_i^{off}$ are numbers of on and off hours of $i^{th}$ generator.

- *Ramp up/down constraint*

Any increasing or decreasing in power output of $i^{th}$ generator for two consecutive periods of time must be limited by ramp up and ramps down, respectively, as follows:

$$P_{i,t} - P_{i,t-1} \leq UR_i \qquad (14)$$
$$P_{i,t-1} - P_{i,t} \leq UD_i \qquad (15)$$

Where $P_{i,t}$ and $P_{i,t-1}$ are power output of $i^{th}$ generator at $t^{th}$ and $t$-$1^{th}$ time, respectively. $UR_i$ and $UD_i$ are up/down ramp rates of $i^{th}$ generator, respectively.

## 2. Uncertainty modeling

As mentioned before, the penetration of RESs into MG can influence the scheduling and operation of MG. PVs and WTs are one of the common types of RESs in active distribution networks and MGs. The generated power by PVs and WTs are coming from the solar irradiation and wind speed as a prime energy source, respectively [36]. Due to the probabilistic nature of wind speed and sun irradiance, the generated power of those resources causes a significant amount of uncertainties. Furthermore, daily load behavior is considered as an uncertain parameter. Therefore, the proposed optimal profit maximization of MG scheduling consists of a large number of uncertain parameters. The probabilistic analysis at the presence of multiple uncertain parameters is a mighty tool for scheduling and operation of power network. To address the uncertain parameters, a probabilistic scenario-based framework is presented in this section.

**2-1. Scenario generation**

*A. WT power output*

The power output of WT depends on the speed of wind. To model the uncertainty of wind speed, It is assumed that the wind speed follows Weibull distribution [7]. If $V_{mean}$ and $\sigma$ are mean and standard deviation of forecasted wind speed, respectively, the parameters of Weibull distribution are calculated as [37]:

$$r = \left(\frac{\sigma}{V_{mean}}\right)^{-1.086} \qquad c = \frac{V_{mean}}{Gamma(1+\frac{1}{r})} \qquad (15)$$

According to the Weibull parameters ($r,c$), the Weibull probability distribution function (PDF) is calculated as:

$$f(V) = \frac{r}{c}\left(\frac{V}{c}\right)^{r-1} \exp\left[-\left(\frac{V}{c}\right)^r\right] \qquad (16)$$

The MCS generates a high number of scenarios subject to the Weibull distribution which each of them is assigned a probability that is equal to one divided by the number of generated scenarios [38]. In each scenario, a random value for the wind speed is considered for the current hour. According to the assigned PDF, in each scenario, the hourly random wind speed is generated, therefore, according to the random wind speed, WT power generation is calculated as:

$$P_W(V) = \begin{cases} 0 & 0 \leq V \leq V_{cut-in} \\ (k_1 + k_2 V + k_3 V^2)P_{W_{rated}} & V_{cut-in} \leq V \leq V_{rated} \\ P_{W_{rated}} & V_{rated} \leq V \leq V_{cut-out} \\ 0 & V_{cut-out} \leq V \end{cases} \qquad (17)$$

where $k_1$, $k_2$ and $k_3$ are coefficients of wind turbine, $P_{W_{rated}}$ is rated power output of WT, $V_{cut-in}$ and $V_{cut-out}$ are minimum and maximum allowable wind speed and $V_{rated}$ is rated wind speed.

*B. PV power output*

The generated power of PV depends on air temperature and solar radiation. To model the uncertainty of irradiation and air temperature, it is assumed that those parameters are following the normal distribution. If $\mu$ and $\sigma$ are mean and standard deviation of forecasted irradiation (air temperature), respectively then the normal distribution PDF for irradiation ($G_{ING}$) and air temperature ($T_r$) is calculated as:

$$f(G_{ING}, T_r) = \frac{1}{\sqrt{2\pi} \times \sigma} \exp\left(-\frac{((G_{ING}, T_r) - \mu)^2}{2 \times \sigma^2}\right) \tag{18}$$

The same as wind speed simulation, a large number of scenarios are generated by MCS to model the PV unit outputs.

According to the assigned PDF, in each scenario, hourly random air temperature and irradiation are generated. Therefore, PV power generation output is calculated as:

$$P_{pv} = P_{STC} \times \frac{G_{ING}}{G_{STC}} \times (1 + K(T_c - T_r)) \tag{19}$$

where $G_{ING}$ is hourly irradiation, $G_{STC}$ is standard irradiation (1000 W/$m^2$), $T_c$ and $T_r$ are cell and air temperature, $P_{STC}$ is rated power of PV and $K$ is the maximum power temperature coefficient [39].

*C. Demand variation modeling*

Due to the load variation during the day, the probabilistic behavior of load is considered as the uncertain parameter. The uncertainty of load demand is subject to normal distribution [4] like (18). Therefore, MCS generates a large number of scenarios. As previous uncertain parameters, in each scenario, hourly random load demand is generated according to the assigned PDF.

**2-2. Scenario reduction**

Initially, a large number of scenarios are generated by MCS. To simplify the computation requirements, the generated scenarios should be reduced. Some of the different scenario reduction techniques are presented in [19, 40]. In this paper, the fast forward selection algorithm is used. The base of this method is to calculate the distance between the scenarios, therefore, the most possible scenarios with more probability are selected. The fast forward selection algorithm works as the following steps [19]:

*Step 1*: Consider $\Omega$ as the initial set of the scenarios: $\Omega = \{1,...\omega_1, \omega_2,....\omega',....N_\Omega\}$. Compute the cost function $\upsilon(\omega, \omega')$ for each pair of scenarios $\omega$ and $\omega'$ in $\Omega$. For example, two simulated wind speed corresponding to $\omega^{th}$ and $\omega'^{th}$ scenarios are 15 and 10 $m/s$, respectively, therefore, the cost function ($\upsilon$) for these two scenarios is: $\upsilon(\omega, \omega') = 15 - 10 = 5$.

*Step 2*: Compute the distance between each pair of the scenarios as follows:

$$d_\omega = \sum_{\substack{\omega'=1 \\ \omega' \neq \omega}}^{N_\Omega} \pi_{\omega'} \upsilon(\omega, \omega') \qquad \forall \omega \in \Omega \qquad (20)$$

where $\pi_{\omega'}$ is probability of $\omega'$th scenario. The scenario with minimum $d_\omega$ is selected (for example $\omega_1$) and $\Omega_s^{[1]} = \{\omega_1\}$. $\Omega_s^{[1]}$ demonstrates the new set of the most probable scenario in the first iteration. When $\omega_1$ is selected, $\Omega_j^{[1]}$ is defined. $\Omega_j^{[1]}$ is equal to the initial set of the scenario ($\Omega$) expect $\omega_1$, therefore: $\Omega_j^{[1]} = \{1, 2,...N_\omega\} \setminus \omega_1$.

*Step 3*: Compute $\upsilon(\omega, \omega')$ for new set of scenarios as:

$$\upsilon^{[2]}(\omega, \omega') = \min\{\upsilon^{[i-1]}(\omega, \omega'), \upsilon^{[i-1]}(\omega, \omega_{i-1})\} \qquad \forall \omega, \omega' \in \Omega_j^{[1]} \qquad (21)$$

According to $\upsilon^{[2]}$, the distance between each pair of scenarios is computed as (20). Like step 2, the scenario with minimum $d_\omega$ is selected (for example $\omega_2$), therefore $\Omega_s$ and $\Omega_j$ are updated as:

$$\begin{aligned} \Omega_s^{[2]} &= \{\omega_1, \omega_2\} \\ \Omega_s^{[2]} &= \{1, 2,...N_\omega\} \setminus \omega_1, \omega_2 \end{aligned} \qquad (22)$$

*Step 4*: Repeat step 2 and 3 until $N_{\Omega_s}$ (number of scenarios in $\Omega_s^{th}$ set) is equal to desired number of scenarios.

*Step 5*: If $\Omega_s^*$ and $\Omega_j^*$ are final sets of selected and deleted scenarios, respectively, and $\omega \in \Omega_s^*$, calculate the probabilities of selected scenarios as:

$$\pi_\omega^* = \pi_\omega + \sum_{\omega' \in J(\omega)} \pi_{\omega'} \tag{23}$$

Where $J(\omega)$ is defined as the set of scenarios $\omega' \in \Omega_j^*$ such that $\omega \in \arg \min_{\omega'' \in \Omega_s^*} \upsilon(\omega'', \omega')$.

## V. Case study

The proposed microgrid problem which is presented in the previous section is implemented on 33-bus microgrid (Fig. 1). The characteristic of three generators is given in Table II. Table III, IV and V show the characteristic of PV, wind and energy storage system, respectively.

Table II : The characteristic of generators of system

| Generator | G 1 | G 2 | G 3 |
|---|---|---|---|
| $P_{min}(kW)$ | 25 | 75 | 25 |
| $P_{max}(kW)$ | 300 | 150 | 300 |
| a($) | 25 | 10 | 20 |
| b($/kW) | 0.15 | 0.85 | 0.25 |
| c($/kW)² | 0.0023 | 0.012 | 0.003 |
| Startup/Shutdown cost ($) | 0.96 | 1.9 | 0.96 |

PV characteristics : Table III

| Technology | $T_c$ | $P_{STC}$ | $G_{STC}$ | $K$ |
|---|---|---|---|---|
| **PV** | $25^0 c$ | 250 kW | $1000\ W/m^2$ | 0.001 |

**Table IV: Wind turbine characteristic**

| | $V_{cut-in}(m/s)$ | $V_{cut-out}(m/s)$ | $V_{rated}(m/s)$ | $P_{wind}^{min}(kW)$ | $P_{wind}^{max}(kW)$ | $k_1$ | $k_2$ | $k_3$ |
|---|---|---|---|---|---|---|---|---|
| **WT** | 3 | 25 | 12 | 0 | 100 | 0.123 | -0.096 | 0.0184 |

**Table V: Energy storage system characteristic**

| ESS | Min/ max charge/ discharge power (kW) | Minimum charge/discharge time (h) | Capacity (kWh) |
|---|---|---|---|
| | -50/+50 | 2 | 100 |

Based on uncertainty modeling for different parameters (PV and wind power generation as well as load demand), which is described in section 2-1, the forecasted value of these parameters are depicted in Fig. 2, 3 and 4.

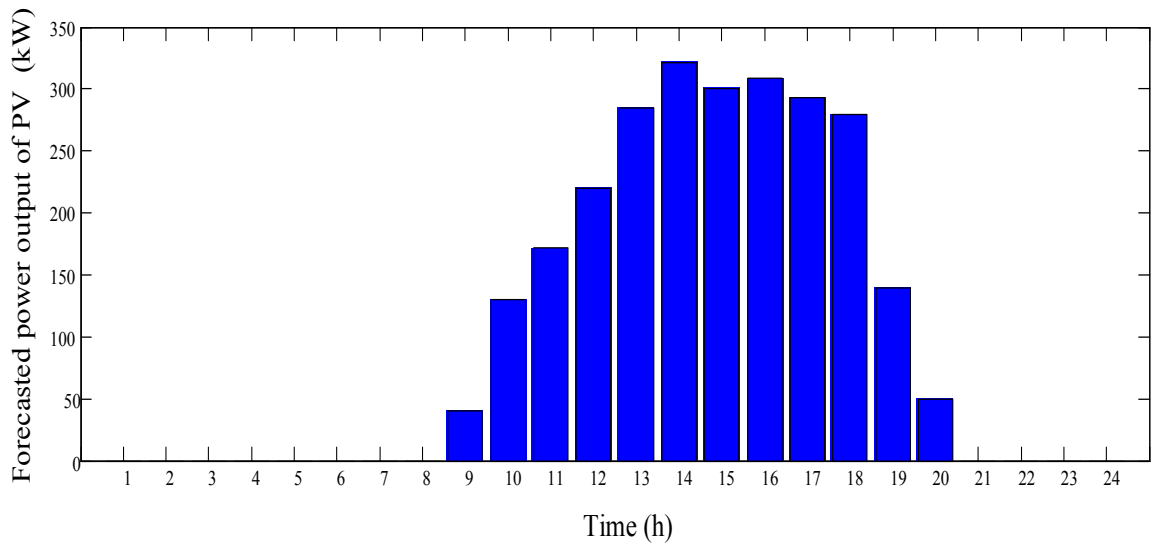

**Fig. 2. The forecasted value of PV power generation**

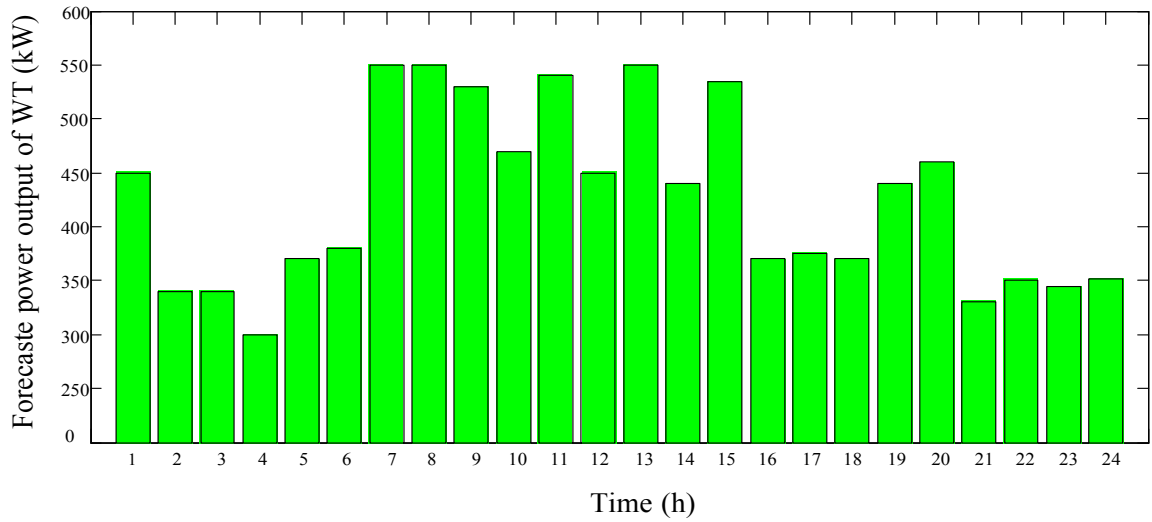

**Fig. 3. The forecasted value of WT power generation**

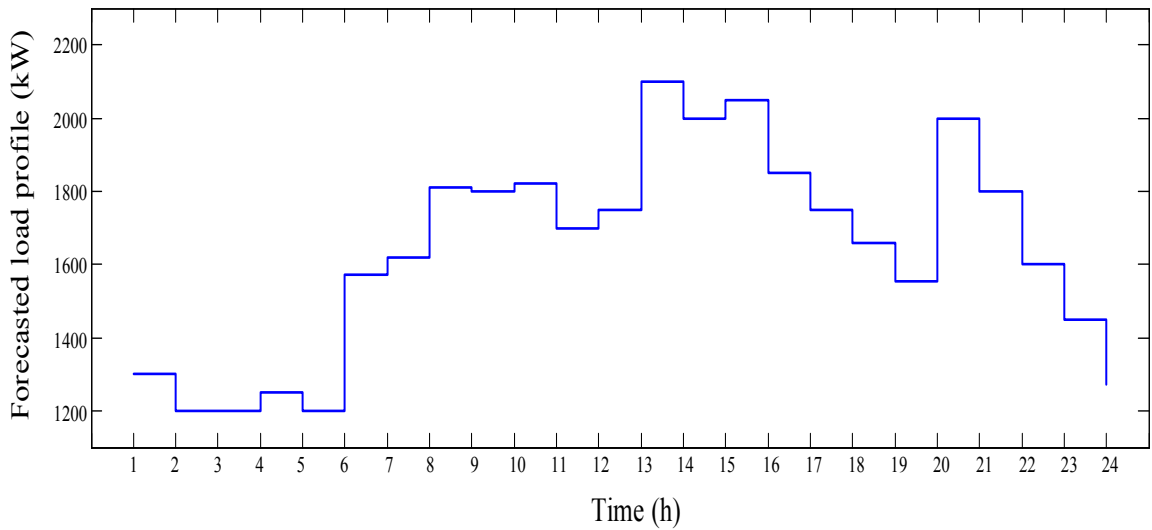

**Fig. 4. The forecasted value of load demand**

The stochastic framework models the power output of WT and PV as well as the load consumption using the corresponding PDF which are described in Section 2. To model the uncertainties, 1000 scenarios are generated for each variable which is reduced to 10 scenarios using scenario generation algorithm (Section 2-2).

Table VI shows the power market and power exchanged prices for each hour. The value of emission factor and emission cost are fixed at 0.003 *kg/kWh* and 0.02 *$/kg*, respectively [33].

Table VI: Hourly energy price [41].

| Hour | Power market price ($/kwh) | Power exchanged price ($/kwh) |
|------|----------------------------|-------------------------------|
| 1    | 0.6   | 1.1  |
| 2    | 0.6   | 1.1  |
| 3    | 0.6   | 1.1  |
| 4    | 0.6   | 1.1  |
| 5    | 0.6   | 1.1  |
| 6    | 0.6   | 1.1  |
| 7    | 0.6   | 1.1  |
| 8    | 0.6   | 1.1  |
| 9    | 0.9   | 1.3  |
| 10   | 0.9   | 1.3  |
| 11   | 0.9   | 1.3  |
| 12   | 1.2   | 1.4  |
| 13   | 1.45  | 1.7  |
| 14   | 1.6   | 1.7  |
| 15   | 1.7   | 1.95 |
| 16   | 1.75  | 1.8  |
| 17   | 1.7   | 1.8  |
| 18   | 1.4   | 1.6  |
| 19   | 1     | 1.3  |
| 20   | 0.8   | 1.3  |
| 21   | 0.8   | 1.25 |
| 22   | 0.8   | 1.3  |
| 23   | 0.7   | 1.2  |
| 24   | 0.6   | 1.1  |

The optimal scheduling of microgrid problem under high level of uncertainties for profit maximization will be solved by the time-varying acceleration coefficient particle swarm optimization (TVAC-PSO) algorithm. It has been discovered by that parameter adapting is a key factor in PSO to find an accurate solution [42]. In TVAC-PSO algorithm, unlike the conventional PSO which considered the fixed coefficient, the acceleration coefficients changed and updated in the search proceeds [43]. So, unlike conventional PSO, the acceleration coefficients updated. More details can be found in [32, 43]. All computer simulations and required coding are carried out in MATLAB software and using CPLEX 11.2 solver.

*Simulation and results*

In this section, the optimal scheduling of MG for profit maximization is analyzed. Based on daily power market price and exchanged power price that is shown the proposed algorithm finds the decision variables $(X_{i,t}, X_{e,t})$. To observe the impact of the proposed scheduling, we execute microgrid optimal scheduling for the reduced scenarios (10 scenarios) and describe one of them with details (power dispatch and hourly cost). Fig. 5 shows the MG profit for 10 scenarios. The scenario number 7 with maximum profit is selected and described in following.

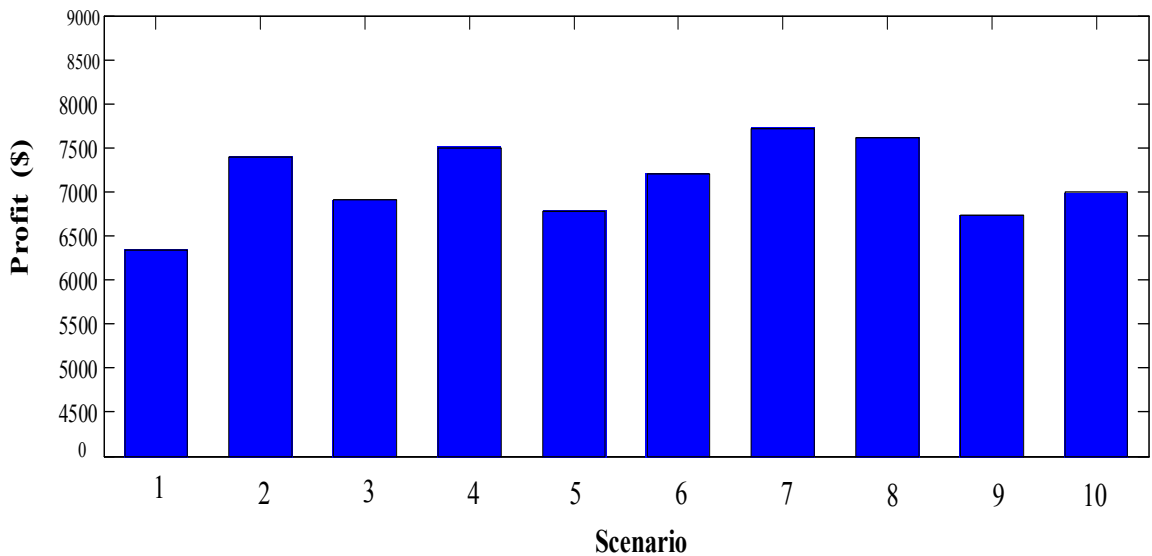

Fig. 5. MG profit for different scenarios.

The optimal generators and ESSs state for 24-h are given in Table VII. ESS charging, discharging and idle states are represented by -1, 1 and 0. As can be seen from Table II, generator number 3 is high-cost generator, so it is committed in peak-load hours (13- to 16-hour).

**Table VII: Results of the 3-generators and ESS states based on decision variable for scenario number 7**

| H | 1 | 2 | 3 | 4 | 5 | 6 | 7 | 8 | 9 | 10 | 11 | 12 | 13 | 14 | 15 | 16 | 17 | 18 | 19 | 20 | 21 | 22 | 23 | 24 |
|---|---|---|---|---|---|---|---|---|---|----|----|----|----|----|----|----|----|----|----|----|----|----|----|----|
| G1 | 1 | 1 | 1 | 1 | 1 | 1 | 1 | 1 | 1 | 1 | 1 | 1 | 1 | 1 | 1 | 1 | 1 | 1 | 1 | 1 | 1 | 1 | 1 | 1 |
| G2 | 0 | 0 | 0 | 0 | 0 | 0 | 0 | 0 | 0 | 0 | 0 | 0 | 1 | 1 | 1 | 1 | 1 | 0 | 0 | 0 | 0 | 0 | 0 | 0 |
| G3 | 1 | 1 | 1 | 1 | 1 | 1 | 1 | 1 | 1 | 1 | 1 | 1 | 1 | 1 | 1 | 1 | 1 | 1 | 1 | 1 | 1 | 1 | 1 | 1 |
| ESS | -1 | -1 | -1 | -1 | -1 | -1 | 0 | 0 | 0 | 0 | 1 | 1 | 1 | 1 | 0 | 0 | 0 | 0 | 1 | 1 | 1 | 0 | 0 | 0 |

The proposed algorithm detects that price of exchanged power is low at some hours (1- to 8-hour), so purchasing power from the upstream network is rational for MG's owner ($r=0$). In this interval, most of ESS is charged. At the higher exchanged power price (10- to 16-hour), it is beneficial for MG to sell surplus power to the upstream network and significant benefits will be achieved. In this interval, ESS is discharged and because the exchanged power price exceeds the cost coefficients of all generators, generators number 2 (high cost generator) committed at 13-hour to 16-hour. While MG supplied their loads, surplus power sold to the upstream network ($r=1$).

The monetary result of the problem consists of MG's costs, revenues and profits are given in Table VIII.

Table VIII: Dispatch schedule and monetary value for scenario number 7

| Hour | Power dispatch (kW) | | | Monetary value | | | | |
|---|---|---|---|---|---|---|---|---|
| | G1 | G2 | G3 | Gen cost($) | Emission cost($) | Energy import cost($) | Total revenue($) | Profit |
| 1 | 300 | 0 | 250 | 547 | 79.2 | 246 | 930 | 57.8 |
| 2 | 300 | 0 | 250 | 547 | 79.2 | 252 | 1005 | 126.8 |
| 3 | 300 | 0 | 250 | 547 | 79.2 | 252 | 1005 | 126.8 |
| 4 | 300 | 0 | 250 | 547 | 79.2 | 298.8 | 942 | 17 |
| 5 | 300 | 0 | 250 | 547 | 79.2 | 241.8 | 1320 | 452 |
| 6 | 300 | 0 | 250 | 547 | 79.2 | 465 | 1038 | --- |
| 7 | 300 | 0 | 250 | 547 | 79.2 | 357 | 1082 | 98.8 |
| 8 | 300 | 0 | 250 | 547 | 100.8 | 540 | 1359.5 | 171.7 |
| 9 | 300 | 0 | 300 | 642 | 115.2 | 415.8 | 1340 | 167 |
| 10 | 300 | 0 | 300 | 642 | 115.2 | 234 | 1379 | 387.8 |
| 11 | 300 | 0 | 300 | 642 | 115.2 | 0 | 1560 | 802.8 |
| 12 | 300 | 0 | 300 | 642 | 115.2 | 0 | 1304.15 | 546.95 |
| 13 | 300 | 100 | 300 | 857 | 194.4 | 0 | 1970 | 918.6 |
| 14 | 300 | 150 | 300 | 1049.5 | 219.6 | 0 | 1600.4 | 331.3 |
| 15 | 300 | 150 | 300 | 1049.5 | 219.6 | 0 | 1433.75 | 164.65 |
| 16 | 300 | 75 | 300 | 783.25 | 190.8 | 0 | 1797.5 | 823.45 |
| 17 | 300 | 0 | 300 | 642 | 115.2 | 0 | 1442 | 684.8 |
| 18 | 300 | 0 | 300 | 642 | 115.2 | 83 | 1145 | 304.8 |
| 19 | 300 | 0 | 300 | 642 | 115.2 | 66.4 | 1015 | 191.4 |
| 20 | 300 | 0 | 300 | 642 | 115.2 | 238.4 | 1500 | 504.4 |
| 21 | 300 | 0 | 300 | 642 | 115.2 | 352 | 1340 | 230.8 |
| 22 | 300 | 0 | 300 | 642 | 115.2 | 294 | 1220 | 168.8 |
| 23 | 300 | 0 | 250 | 547 | 79.2 | 322.5 | 1095 | 146.3 |
| 24 | 300 | 0 | 250 | 547 | 79.2 | 210 | 1186 | 349.8 |
| Total | | | | Total cost ($) 23287.95 | | | 31009.3 | 7721.35 |

It is seen from Table VIII, at the peak load hours (11-16) which the value of power price exceeds generators cost coefficients, MG sells the surplus power (that generated by high-cost generator) to the upstream network and receives significant revenues, in this interval, $r$ that shows the exchanged power direction with the upstream network is equal 1. Based on the proposed optimal scheduling, the profit of MG is determined as $7721.35 for scenario number 7.

Finally, the expected profit of 10 reduced scenarios based on the probability of each scenario can be calculated as follows:

$$\textbf{Expected Profit} = \sum_{\omega=1}^{10} \pi_\omega f_\omega \quad \text{where: } \sum_{\omega=1}^{10} \pi_\omega = 1 \tag{24}$$

According to the (24), the value of expected profit is **$ 7528.9**.

## VI. Conclusion

In this chapter, a brief overview of uncertainty management in the modern power system is studied. The penetration of renewable energy resources, as well as load demand deviation, cause different challenges in the operation and scheduling of modern power systems. After classifying a variety of uncertain parameters in the power system, some useful method for uncertainty management is discussed. To demonstrate the uncertainty modeling in power system decision-making process, we considered a microgrid consists of different generation units (dispatchable units and renewable units like PV and wind). The MG scheduling for profit maximization in the presence of multiple uncertain parameters is examined. After the formulation of the problem, generated power by renewable energy and load demand are considered as the uncertain parameters and stochastic scenario-based approach is used for solving the problem. The proposed method was examined on the 33-bus microgrid test system. The result of decision variables of the problem consists of the status of the storage and dispatchable units as well as power output of dispatchable units, exchanged power with the upstream network, and the monetary value was studied for the best scenario. Finally, the expected profit of reduced scenarios was calculated. The results show that by modeling the uncertainty in a lot of scenarios, the operator of the system can decide with a better view, about the conditions of the network. However, the high volume of computations in this method requires an examination of other methods which will be studied in future works.

## Acknowledgment

The work done by Alireza Soroudi is supported by a research grant from Science Foundation Ireland (SFI) under the SFI Strategic Partnership Programme Grant No. SFI/15/SPP/E3125. The opinions, findings and conclusions or recommendations expressed in this material are those of the author(s) and do not necessarily reflect the views of the Science Foundation Ireland.